\documentclass[preprint,aps,prd,amsmath,amssymb]{revtex4}
\usepackage{graphicx}
\usepackage{color}
\usepackage[latin1]{inputenc}
\newcommand{\be}{\begin{eqnarray}}
\newcommand{\beq}{\begin{equation}}
\newcommand{\eeq}{\end{equation}}
\newcommand{\ee}{\end{eqnarray}}

\newcommand{\bmp}{\noindent\begin{minipage}{16cm}}
\newcommand{\emp}{\end{minipage}\vskip 7mm} 

\usepackage{graphicx}
\usepackage{dcolumn}
\usepackage{bm}
\usepackage{amsmath}
\usepackage{amsfonts}
\usepackage{bbm}
\usepackage{subfigure}
\usepackage{feynmf}
\usepackage{latexsym}
\usepackage{amssymb}

\def\drawbox#1#2{\hrule height#2pt
        \hbox{\vrule width#2pt height#1pt \kern#1pt
              \vrule width#2pt}
              \hrule height#2pt}

\def\Asym#1#2{\vcenter{\vbox{\drawbox{#1}{#2}
              \kern-#2pt 
              \drawbox{#1}{#2}}}}

\begin{document}
\title{Corrigan-Ramond Extension of QCD at Nonzero Baryon Density}
\author{M.T. {\sc Frandsen}}
\email{toudal@nbi.dk}
\author{C. {\sc Kouvaris}}
\email{kouvaris@nbi.dk}
\author{F. {\sc Sannino}}
 \email{francesco.sannino@nbi.dk}
\affiliation{The Niels Bohr Institute, Blegdamsvej 17, DK-2100 Copenhagen \O,
 Denmark }


\begin{abstract}

We investigate the Corrigan-Ramond extension of one massless flavor Quantum Chromo Dynamics at nonzero quark chemical potential.
Since the extension requires the fermions to transform in the two index antisymmetric representation of the gauge group, one finds that the number of possible channels is richer than in the 't Hooft limit. We first discuss the diquark channels and show that for a number of colors larger than three a new diquark channel appears. We then study the infinite number of color limit and show that the Fermi surface is unstable to the formation of the Deryagin-Grigoriev-Rubakov chiral waves. We discover, differently from the 't Hooft limit, the possibility of a colored chiral wave breaking the color symmetry as well as
translation invariance.

\end{abstract}


\maketitle

\section{Introduction}

Different limits have been proposed
in order to get a better understanding of the complex structure
of strongly coupled gauge dynamics and more specifically of
 Quantum Chromo Dynamics (QCD).
At zero temperature and baryon density, one of the best
 known limits is the large number of colors N proposed by 't Hooft
\cite{'tHooft:1973jz}. Here, one increases the number of colors while
the
 quarks transform in the
fundamental representation of $SU(N)$. It is then possible to organize a diagrammatic expansion in the inverse
of the number of colors. The hope is that the leading terms in the large N expansion may provide a good description of the
three color physical world.
 Indeed a number of QCD properties have a simple
 understanding in this limit
\cite{Witten:1979kh}. Recently, also modern lattice simulations
 have explored the
't Hooft limit \cite{Bringoltz:2005rr,Bringoltz:2005az,Narayanan:2005gh}.
However, there are cases in which the leading terms in the 't Hooft expansion are not sufficient to capture
basic properties of QCD. For example it has been
 recently shown, that the
 low energy meson-meson scattering
amplitudes \cite{Harada:2003em} are not  well represented by the 't Hooft large N limit.
Hence, it is interesting to investigate different limits which
 for certain values of the theory parameters also yield QCD.

Soon after the 't Hooft limit, Corrigan and Ramond (CR) proposed an
 alternative QCD extension \cite{Corrigan:1979xf}. Here
 the quarks transform, with respect to the
gauge group, according to the two index antisymmetric
representation. To be more
precise Corrigan and Ramond \cite{Corrigan:1979xf} suggested a
generalization of QCD in which some flavors were in a higher
dimensional representation and others still in the fundamental
representation. The focus there was on the baryonic and the $\eta^{\prime}$ properties of QCD. It is easy to show that for three colors
 the CR limit is identical to QCD. However at
large number of colors the CR and 't Hooft limits are very different. For instance
 quark loops are suppressed in the 't Hooft limit
but not in the CR extension of QCD. Some of the properties at leading number of colors and the possible relevance for QCD were suggested already in the
Corrigan and Ramond work  \cite{Corrigan:1979xf} and further investigated
 by Kiritsis and Papavassiliou \cite{Kiritsis:1989ge}.

Recently, Armoni, Shifman and Veneziano \cite{Armoni:2003gp} have also proposed an interesting relation between certain sectors of the two index antisymmetric (and symmetric) theories at large number of colors and
sectors of super Yang-Mills (SYM). This has lead to a renewed interest in this limit. Using a supersymmetric inspired
effective Lagrangian approach some of the $1/N$ corrections were
investigated in \cite{hep-th/0309252}. Shifman and one of the present authors derived interesting consequences about the spectrum and the vacuum of QCD and also of non-QCD like theories. {}From this work one can start understanding the mechanism which makes the scalar companions of the lowest-lying pseudoscalars heavy with respect to their chiral partners. It is important to note that despite the enormous amount of theoretical information about supersymmetric gauge theories we still do not have knowledge of basic properties. For example, one cannot even answer the simple question of which hadronic states are the lightest ones in SYM. In \cite{Merlatti:2004df} it was
 shown that the lightest states are constituted by the supermutliplet containing the
gluinoball while the supemutiplet of glueballs is heavier. If supersymmetry will not be observed in experiments this might be the only way we can infer some {\it experimental} information on these theoretically relevant theories. Different sectors in the CR large N limit are, however, not mapped in SYM.

Besides these two limits a third one for massless one-flavor QCD, which is, somewhat, in between the 't Hooft and the CR
one, has been proposed very recently \cite{Ryttov:2005na}. Here, one first splits the QCD Dirac fermion into the two elementary Weyl fermions and afterwards assigns one of them to transform according to a rank-two antisymmetric tensor while the other remains in the fundamental representation of the gauge group. {}For three colors one reproduces
one-flavor QCD and for a generic number of colors the
theory is chiral. Such a theory turns out to be a particular case of the generalized
Georgi-Glashow (gGG) model.

{}All of the previous very encouraging results suggest that it is relevant to explore, within the CR extension of QCD, other regimes.
At nonzero temperature, for example, the confinement/deconfinement phase transition problem was studied in \cite{hep-th/0507251}. In this paper it was shown that an alternating pattern, as a function of number of
colors, with respect to the symmetries of the center group appears. The confinement/deconfinement properties in the
CR limit are expected to be different than the SYM ones \cite{hep-th/0507251}. It has also been very instructive to investigate the relation between
confinement and chiral symmetry in the CR extension of QCD and then confront them with the 't Hooft one \cite{hep-th/0507251}.

In the present work we provide the first investigation of the CR extension of QCD when
turning on a nonzero baryon density.  Much recent work has been
devoted to try to unveil possible phases of QCD at nonzero baryon
density. It has been established, using various techniques, that
the ground state of QCD with two or three flavors at large quark
chemical potential displays a color superconductive phase
\cite{Barrois:1977xd,Bailin:1983bm,Alford:1997zt,Rapp:1997zu}. One
can also determine the pattern of chiral symmetry breaking
 at asymptotically high density and show, for example, that for three degenerate and light flavors
the Color-Flavor-Locked phase (CFL) \cite{Alford:1998mk} emerges.

In the hope to gain some further insight, the large N 't Hooft limit at nonzero baryon density has also been investigated, first in \cite{Deryagin:1992rw}
 and more recently in \cite{Shuster:1999tn}. One discovers that color superconductivity
is suppressed in the 't Hooft large N limit and that the Fermi
 surfaces are unstable to chiral waves. These waves correspond to
chiral condensates that vary as a function of the space
 coordinates and therefore break translational invariance.

 Since the CR extension requires the fermions to transform according to the two index antisymmetric representation of the gauge group we find a richer group theoretical structure than in the 't Hooft limit. We first, briefly, discuss the diquark channels and show that when N is larger than three a new diquark channel appears. Another interesting feature is that for an even number of colors the center group symmetry breaks to a $Z_2$ and hence we expect these theories to display also color confinement \cite{hep-th/0507251}. We then study the infinite number of color limit and show that, as for the 't Hooft case, color superconductivity is not favored with respect to the formation of the Deryagin-Grigoriev-Rubakov (DGR) chiral waves. We discover, differently from the 't Hooft limit, the possibility of the  presence of a colored chiral wave breaking the color symmetry as well as
translation invariance. The presence of the DGR instability in the CR limit of QCD shows that, even at an infinite number of colors the baryonic sector of the theory does not decouple at nonzero matter density.

\section{Notation and conventions}
It is useful to
formulate in detail the theory for quarks in the antisymmetric representation of the gauge group. The Lagrangian reads:
\begin{eqnarray}
\mathcal{L}  =  -\frac{1}{2}\text{Tr}\left[F^{\mu\nu}F_{\mu\nu}\right]
+ i\bar{q}\gamma^{\mu}
D_{\mu}q\nonumber
 + \mu \, {q}^{\dagger} q,
\end{eqnarray}
with the Dirac fermion $q$ :
\begin{eqnarray}
q^{[ij]} & = &  q^{\tilde{a}}(t^{\tilde{a}})^{ij} \ , \qquad \tilde{a}=1,\ldots ,\frac{N(N-1)}{2} \ ,
\end{eqnarray}
and $i,j=1,\ldots, N$. The generators, in the fundamental representation of the $SU(N)$ gauge group are the matrices $t^a$ with $a=1,\ldots, N^2-1$ normalized
according to $\rm{Tr}\left[t^a t^b\right]=\delta^{ab}/2$ and $t^{\tilde{a}}$ are the subset of these matrices which are antisymmetric.
The gauge field and the field strength are defined as:
\begin{eqnarray}
A_{\mu}  =  A_{\mu}^a t^a \ ,
\qquad
F_{\mu\nu}  =  \partial_{\mu}A_{\nu}-\partial_{\nu}A_{\mu} + i g
[A_{\mu},A_{\nu}] \ .
\end{eqnarray}
The covariant derivative acting on the fermionic fields is:
\begin{eqnarray}
D_{\mu}q^{\tilde{a}} & = & \partial_{\mu} q^{\tilde{a}}+ i g A_{\mu}^{a}(T^a)^{\tilde{a}}_{\tilde{b}} q^{\tilde{b}} \ ,
\end{eqnarray}
with the generator $T^a$ written explicitly for the two index antisymmetric representation via the generators in the fundamental representation:
\begin{eqnarray}
(T^a)^{\tilde{a}}_{\tilde{b}}= 4{\rm Tr} \left[t^{\tilde{a}} t^a t^{\tilde{b}}\right] \ , \quad {\rm with} \quad
{\rm Tr} \left[T^a T^b\right] = \frac{N-2}{2} \,\delta^{ab} \ .
\end{eqnarray}
Of course one can also define the generators of the two index antisymmetric representation directly as tensor product of the generators in the fundamental representation. We have also introduced directly the chemical potential for the baryon number of the theory and limit ourself to the one flavor case. It is also easy to check that for three colors the previous Lagrangian describes one massless flavor QCD.

\section{CR and the DGR Instability}

We start our preliminary study of the CR extension of QCD at nonzero quark chemical
potential. We consider the limit in which the baryon density is large enough that perturbative techniques are
applicable. Throughout the work we assume the fermions to be massless and identify the Fermi energy and momentum (e.g. $p_F=\mu$). Some of the key properties of the large chemical potential limit
 can be explored using renormalization group techniques.
At small temperatures the presence of a large baryon chemical potential induces the formation of a Fermi surface of quarks. Renormalization group techniques allows us to concentrate on the relevant scattering processes. The large number of colors limit together with a nonzero baryon chemical potential in the 't Hooft limit has been investigated in
\cite{Deryagin:1992rw,Shuster:1999tn}, here we will extend this analysis to the CR limit.

Cooper pair formation is favored by any attractive interaction near the Fermi surface.
This fact leads to the phenomenon of superconductivity. {}For QCD at high chemical potential the
Cooper pairs carry color charges and hence generically color symmetry breaks spontaneously.

The formation of a Cooper pair is often described as due to the presence of an instability
near the Fermi surface, i.e. the BCS instability. In general there will be more than one  attractive interaction in the system. Each different attractive interaction tends to form Cooper pairs with different quantum numbers. Renormalization Group (RG) analysis near the Fermi surface can be used to determine which attraction dominates.

A practical procedure to uncover the presence of an instability in a given channel is to determine first the
low energy effective coupling constant of the associated 4-fermion interaction near the Fermi surface. This effective coupling constant contains
color, flavor and spin dependence.
If the effective coupling constant develops a Landau pole, this indicates an
instability in the associated channel. It is expected that the channel in which the Landau pole is reached first is the one which occurs. A review of the RG analysis for BCS theories can be found in \cite{Shankar:1993pf} and was introduced for investigating color superconductivity in \cite{Evans:1998ek}. The situation at asymptotically large chemical potentials where perturbation theory is valid is more involved. As Son \cite{Son:1998uk} has demonstrated magnetic gluons are not screened at large quark chemical potential which means that the one gluon exchange does not reduce exactly to a contact term interaction.

Given the above we start our analysis of the CR limit by considering quark-quark scattering at the
Fermi surface. Since we are only interested in the large N limit behavior we will not attempt a full analysis of the ground state at a low number of colors. {}For our purposes it will be sufficient to investigate the overall color structure.

\begin{figure}[!htp]
\begin{center}
\includegraphics[scale=0.9]{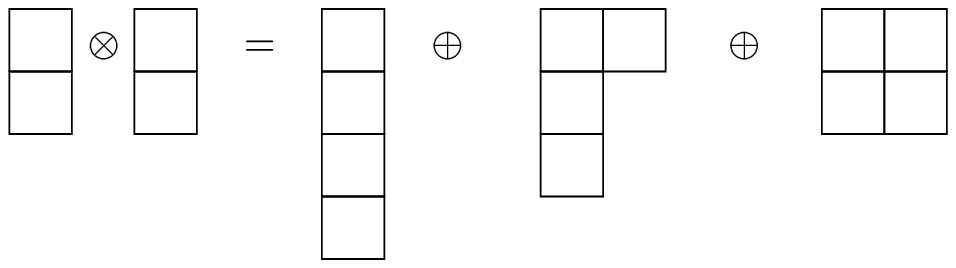}
\caption{Decomposition of the tensor product
 of two quarks in the 2-index antisymmetric representation.
In the text we refer to the three channels that appear in the RHS
as (a) , (b) and (c) respectively.}
\label{young_qq}
\end{center}
\end{figure}

Since our quarks transform according to the
 2-index antisymmetric representation of the
 gauge group, for a generic number of colors $N>3$,
the scattering leads, in color space, to the three channels
depicted in Fig.~\ref{young_qq} via Young-Tableaux and labelled by (a), (b), and (c)
 respectively.

 For $N=3$ the channel (a) on the RHS of Fig.~\ref{young_qq}
disappears and the two channels ((b) and (c)) correspond to the antitriplet and the sextet
  of QCD. The former leads to the instability responsible for color superconductivity in QCD.

Note that when extending the theory to a number of colors larger than $3$ while keeping the fermions in the fundamental representation,
as in the 't Hooft limit, only two channels appear. They are the straightforward generalization of the ones in QCD. However in the CR generalization
 of QCD also the channel (a) appears and it turns out to be the most attractive one.
This is so since it is antisymmetric with respect to all of the color indices. Also channel (b) is attractive but is surely dominant only for $N=3$, i.e. for QCD. We note that for four colors the most attractive channel is a color singlet.

We summarize here the color factors entering the RG analysis for the
quark-quark scattering three channels stemming from the one gluon exchange interaction
 of Fig.~\ref{young_qq}:
 \begin{eqnarray}
C_a = 2 + \frac{2}{N} \ , \qquad C_b = \frac{2}{N} \ , \qquad C_c = - 1 + \frac{2}{N} \ .
\end{eqnarray}
The coupling of the relevant four fermion interaction will be proportional, in each channel, to the respective coefficient above. In our conventions positive coefficients lead to attractive interactions. If the ground state, indeed, corresponds to the case (a) we then also expect the Cooper pair to carry spin one.

 As an aside we note that differently from the case with fermions in the fundamental representation, for an even number of colors the underlying theory still preserves a $Z_2$ center group symmetry. This fact allows one to define the Polyakov loop as a good order parameter for confinement. This is interesting since for an even number of colors we predict the presence of a well defined deconfining phase transition \cite{hep-th/0507251}. It would then be interesting to investigate the relation between the confinement and superconductive phase transition. We expect this case to be similar to the one discussed in \cite{Sannino:2004ix,{Mocsy:2003qw}}.

The large number of  colors is also interesting in the CR limit due to the possibility that,
at least in vacuum, certain sectors of the theory are expected to be mapped in super Yang Mills.
Confinement properties, however, cannot be matched at zero and nonzero temperature  \cite{hep-th/0507251}. Since super Yang Mills does not possess any baryon number it is clear that little information can be deduced using the correspondence when
exploring the presence of a nonzero chemical potential. However the CR limit is still well defined and using the large chemical potential limit
one can derive results which are under perturbative control. One could, however, still hope that at large number of colors the baryon number decouples and the knowledge of the supersymmetric theory be of help. A possible naive argument is that at large number of colors nonplanar diagrams are suppressed and hence diquark condensation is suppressed and one expects a quark-antiquark condensate.

As we shall shortly see superconductivity is indeed suppressed at large N in the CR limit but now the Fermi surface is unstable with respect to the development of  chiral waves with $2\mu$ wave number:
\begin{eqnarray}
\langle \bar{q}(x) q(y)\rangle  = e^{i\vec{P}\cdot {(\vec{x} + \vec{y})}}  \int d^4q \,
e^{-iq\cdot(x-y)} f(q) \ ,
\end{eqnarray}
 where $\vec{P}$ has modulus $P=\mu$ and has arbitrary direction. We will discuss the color structure of the previous condensate in the large CR limit later in the text.

The DGR instability has been first discovered at infinite number of colors in the 't Hooft limit of QCD \cite{Deryagin:1992rw}. Due to the new group structure, the CR large N limit leads to different possibilities in the color composition of the pair with respect to the 't Hooft case.

There is an important difference with respect to the ordinary constant chiral condensate. In vacuum the pairing happens between a particle and an antiparticle moving in opposite directions. In the present case one pairs a particle and a hole near the Fermi surface moving in the same direction and with a momentum near the Fermi one. The scattering is nearly in the forward direction and the scattering amplitude becomes singular favoring the formation of a pair.

\begin{figure}[!htp]
\begin{center}
\includegraphics[scale=.9]{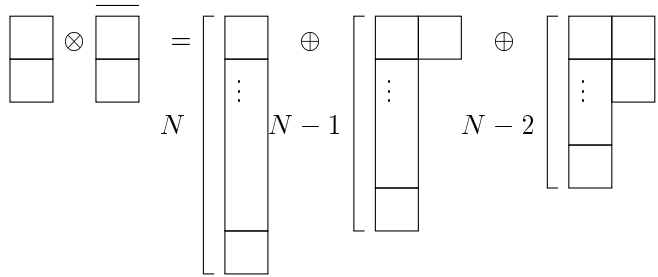}
\caption{Decomposition of the tensor product of a quark in the 2-index
antisymmetric representation and an antiquark in the
anti-2-index antisymmetric one. The three channels in the RHS are
the singlet, the adjoint and an extra channel not present in the 't Hooft limit. }
\label{young_qqb}
\end{center}
\end{figure}

To determine the existence of the DGR instability in the large N CR limit we use again the renormalization group approach and
consider an infinite number of colors \cite{Deryagin:1992rw,Shuster:1999tn}. So we start with the study of $\bar{q}q$ scattering. In
this case we have a quark in the
2-index antisymmetric representation and a hole in the
anti-2-index representation. In Fig.~\ref{young_qqb} we show the
 decomposition of the tensor product of the two representations. In the RHS of Fig.~\ref{young_qqb} we have respectively the
 totally antisymmetric channel (singlet),
 the adjoint and a
third one. The first two channels correspond exactly to the ones in the 't Hooft limit while the third one
is a specific feature of the CR extension of QCD. Note, however, that since the scattering states are in the two index antisymmetric representation
this also affects the overall color factor for the scattering process under consideration so that one cannot simply use the results of the large N a la 't Hooft for the singlet and the adjoint channel.
As we mentioned at the beginning of the section one might imagine the possibility of the formation of a homogeneous condensate $\langle
\bar{q} q \rangle$. In this case the condensate is associated to pairing a quark and an antiquark at large quark matter densities.
However this is not an option since the amount of energy needed in order to make such a pair is
twice the quark chemical potential. Following Deryagin, Grigoriev, and
Rubakov \cite{Deryagin:1992rw}
 and later Shuster and Son \cite{Shuster:1999tn} we consider the possibility to form Cooper pairs among
 a quark and a hole. This pairing is energetically favored with respect to the quark-antiquark one.

Near the Fermi surface we consider a quark with momentum
$P+q$, and the hole with momentum $P-q$. The total momentum of the pair is $2P$ and is approximately equal to twice the Fermi momentum. $q$ is
a small fluctuation of the momentum above the Fermi surface.

Shuster and Son introduced an elegant way to derive the DGR results in the 't Hooft limit. The method consists in reducing
the problem to a 1+1 dimensional one and it was invented in the past in a different context \cite{ILA}. Since the derivation uses only kinematical considerations which can be applied directly to the CR limit we will not repeat them here. We just recall that one can decompose the small momentum $q$ in a component parallel to $P$ and the orthogonal one. One can show that at sufficiently large chemical potential and within a well defined kinematical window one can neglect the orthogonal component.

Remarkably the relevant effective theory in 1+1 dimensions is the following non-abelian Thirring model {\it but} with a
momentum dependent coupling constant \cite{Shuster:1999tn}
\beq
L_{eff}=i \bar{\Psi}\gamma^{\mu}\partial_{\mu}\Psi -
 \frac{g^2}{4\pi}\ln\frac{\Delta}{q_{||}}(\bar{\Psi}\gamma^{\mu}
\frac{T^{a}}{2}\Psi)^2 \ .
\eeq
$\Delta$ denotes the scale for the perpendicular
 momentum $q_{\perp}$ to $P$ and $q_{||}$ is the
longitudinal component of $q$.  The $\gamma_{\mu}$ are two dimensional gamma matrices.
If we choose $P$ to be along the z axis, $\Psi$ is the following field in terms of the original four dimensional Dirac quark:
\beq
\Psi =
\left(
\begin{array}{c}
e^{-i \mu z} q_{L2} \\
e^{i \mu z} q_{R2}  \\
e^{-i \mu z} q_{R1} \\
e^{i \mu z} q_{L1}
\end{array}
\right),
\eeq
where $q_{L2}$, $q_{R2}$, $q_{R1}$, and $q_{L1}$
  are identified as the four components of the
usual Dirac spinor
 $q^T = (q_{L1}\, q_{L2} \, q_{R1} \, q_{R2})$
 in the Weyl basis. We have suppressed the color indices. This particular decomposition becomes more clear when we recognize that we
 are only considering the positive energy particle states (i.e. $q_{L2(L1)}$, $q_{R1(R2)}$) when the particle's momentum is near $\vec{P}(-\vec{P})$.
 The exponential factors are introduced to eliminate the fast spatial variations of the fields.

 Using the above identification, it is easy
 to see that a constant condensate
 of the form $\langle \bar{\Psi} \Psi \rangle$
 corresponds to space-dependent condensates of
$\langle \bar{q} q \rangle$, because
 $\langle \bar{\Psi} \Psi \rangle = \cos (2\mu z) \langle \bar{q} q \rangle
-i\,\sin (2 \mu z)\langle \bar{q}\gamma^0 \gamma^3 q\rangle$. In the RHS of the last expression we are using the four dimensional gamma matrices.

The running of the coupling of the effective theory is governed again by an equation of the type:
\beq
\label{beta}
\frac{\partial \lambda (s)}{\partial s}=2\frac{C}{\pi} \lambda^2 (s) \ .
\eeq
The effective coupling constant depends on $q_{||}$
and consequently $\lambda$ should be a function of the RG parameter s and $q_{||}$.

The relevant kinematical range is for $\Delta^2/\mu <q_{||}< \Delta$. The renormalization group equation needs to be modified \cite{Shuster:1999tn} since the four fermion coupling constant has also a direct dependence on the renormalization scale $s$ via the direct dependence on $q_{||}$.
 We can link $s$ and $q_{||}$ recalling that the typical momentum scale of the internal lines in the loop is of the order of  $\Delta e^{-s}$ which sets also the
vertex energy scale. One can show that the coupling constant hits a Landau pole, and hence signals an instability, when $s_L=\pi/h2$, where $h=2g^2 C /4\pi^2$. This translates via the $\Delta e^{-s}$ to the following Landau energy scale
\beq
\label{pole}
E_L=\Delta e^{-\frac{\pi}{2h}}=\Delta e^{-\frac{\pi^2}{\sqrt{2g^2 C}}} \ .
\eeq
 Before
discussing the results for the CR limit we review the 't Hooft limit ones \cite{Deryagin:1992rw,Shuster:1999tn}. Here one has only  two different channels, the totally
 antisymmetric (singlet) and the one that corresponds to the octet
in QCD. However the factor C is
 proportional to N  only
for the singlet case while it is a constant for the adjoint channel. {}From (\ref{pole})
it is clear that the larger the value of C the
 earlier we meet the Landau pole
 in the RG flow. This is an indication in favor of the chiral wave instability in the singlet channel \cite{Deryagin:1992rw,Shuster:1999tn}.

{}For the CR extension of QCD at large number of colors we find the
following values of the one loop relevant coefficient $C$:
\begin{eqnarray}
C_{\rm singlet}= N \ ,\qquad  C_{\rm Adjoint}= \frac{N}{2} \ , \qquad C_{\rm extra} = - \frac{2}{N} \ .
\end{eqnarray}
The subscript {\it extra} refers to the third channel in the CR extension of QCD which is not present in the 't Hooft limit. The extra channel does not lead to an instability and is subleading in N with respect to the other two channels. Our results show
that both the singlet and the adjoint channel in Fig.~\ref{young_qqb}
are leading in N. This result is somewhat unexpected. We recall that in
the 't Hooft limit the adjoint channel, although attractive, is subleading with respect to the singlet one. The difference in the
large N behavior here is due to the fact that we have many more fermions in the CR limit than in the 't Hooft one.

It is reasonable to expect that for
$g^2N<<1$, see (\ref{pole}), the singlet channel dominates, since there is still a small numerical suppression due to the extra 1/2 factor in the adjoint channel. A situation similar to the 't Hooft limit would then emerge, at least for $N\rightarrow \infty$. However such a small suppression factor does not preclude the possibility, especially when increasing the value of $g^2 N$, that either both channels occur simultaneously or that the adjoint channel is the relevant one. In both cases color symmetries break spontaneously due to the presence of colored chiral waves.

Another remark is that one has shown that the theory is sensitive to the presence of the baryon number even at infinite number of colors. Besides, the condensate which forms has an explicit dependence on space (i.e. is a chiral wave). In this way one has also explicitly tested the baryon number dependence of the CR extension of QCD and shown that the baryon number, at nonzero chemical potential, does not decouple even in the planar limit \cite{Armoni:2003gp}.

\section{Conclusions}
We have investigated the CR extension of one massless flavor QCD at nonzero quark chemical potential. We have shown that this extension, both for a small as well as an infinite number of colors leads to a richer structure than in the 't Hooft case.

We have first discussed the diquark channels and have seen that for a number of colors larger than three a new diquark channel appears.
We have also considered the infinite number of color limit. Here, color superconductivity is not favored with respect to the Deryagin-Grigoriev-Rubakov chiral wave phase. Differently from the 't Hooft limit there is also the possibility of a colored chiral wave which breaks both color symmetry and translation invariance.

One can now extend the work to a finite number of colors \cite{Shuster:1999tn} or to a larger number of flavors. There are a number of possibilities, one can add new flavors in the fundamental (following CR) or in the two index antisymmetric representation. It is also very interesting to investigate in some detail the interplay between confinement and chiral symmetry in this limit at nonzero temperature and quark chemical potential. Our results show that even in the infinite number of colors limit in the CR extension of QCD one is still sensitive to the baryon symmetry.

\acknowledgments
We are very happy to thank S.B. Gudnason and T.A. Ryttov, respectively for helping with the figures and for sharing some of their notes.

The work of C.K. and F.S. is supported by the Marie Curie Excellence Grant under contract MEXT-CT-2004-013510 and F.S. is also supported by the Danish Research Agency.


\begin{thebibliography}{199}


\bibitem{'tHooft:1973jz}
  G.~'t Hooft,
  ``A Planar Diagram Theory For Strong Interactions,''
  Nucl.\ Phys.\ B {\bf 72} (1974) 461.

\bibitem{Witten:1979kh}
  E.~Witten,
  ``Baryons In The 1/N Expansion,''
  Nucl.\ Phys.\ B {\bf 160}, 57 (1979).

\bibitem{Bringoltz:2005rr}
  B.~Bringoltz and M.~Teper,
  ``The pressure of the SU(N) lattice gauge theory at large-N,''
  arXiv:hep-lat/0506034.

\bibitem{Bringoltz:2005az}
  B.~Bringoltz,
  ``The critical region of strong-coupling lattice QCD in different large-N
  limits,''
  arXiv:hep-lat/0511058.

\bibitem{Narayanan:2005gh}
  R.~Narayanan and H.~Neuberger,
  ``The quark mass dependence of the pion mass at infinite N,''
  Phys.\ Lett.\ B {\bf 616}, 76 (2005)
  [arXiv:hep-lat/0503033].


\bibitem{Harada:2003em}
  M.~Harada, F.~Sannino and J.~Schechter,
  ``Large N(c) and chiral dynamics,''
  Phys.\ Rev.\ D {\bf 69}, 034005 (2004)
  [arXiv:hep-ph/0309206].


\bibitem{Corrigan:1979xf}
  E.~Corrigan and P.~Ramond,
  ``A Note On The Quark Content Of Large Color Groups,''
  Phys.\ Lett.\ B {\bf 87}, 73 (1979).

\bibitem{Kiritsis:1989ge}
  E.~B.~Kiritsis and J.~Papavassiliou,
  ``An Alternative Large N Limit For QCD And Its Implications For
Low-Energy Nuclear Phenomena,''
  Phys.\ Rev.\ D {\bf 42}, 4238 (1990).

\bibitem{Armoni:2003gp}
  A.~Armoni, M.~Shifman and G.~Veneziano,
  `Exact results in non-supersymmetric large N orientifold field
theories,''
  Nucl.\ Phys.\ B {\bf 667}, 170 (2003)
  [arXiv:hep-th/0302163].
A.~Armoni, M.~Shifman and G.~Veneziano, ``SUSY relics in one-flavor
QCD from a new 1/N expansion,'' Phys.\ Rev.\ Lett.\ {\bf 91} (2003)
191601, [arXiv:hep-th/0307097].

\bibitem{hep-th/0309252}
  F.~Sannino and M.~Shifman,
  ``Effective Lagrangians for orientifold theories,''
  Phys.\ Rev.\ D {\bf 69}, 125004 (2004)
  [arXiv:hep-th/0309252].

\bibitem{Merlatti:2004df}
  P.~Merlatti and F.~Sannino,
  ``Extending the Veneziano-Yankielowicz effective theory,''
  Phys.\ Rev.\ D {\bf 70}, 065022 (2004)
  [arXiv:hep-th/0404251].
  A.~Feo, P.~Merlatti and F.~Sannino,
  ``Information on the super Yang-Mills spectrum,''
  Phys.\ Rev.\ D {\bf 70}, 096004 (2004)
  [arXiv:hep-th/0408214].

\bibitem{Ryttov:2005na}
  T.~A.~Ryttov and F.~Sannino,
  ``Hidden QCD in chiral gauge theories,''
  arXiv:hep-th/0509130. To appear in Phys. Rev. D.

\bibitem{hep-th/0507251}
  F.~Sannino,
  ``Higher representations: Confinement and large N,''
  arXiv:hep-th/0507251. To appear in Phys. Rev. D.

\bibitem{Barrois:1977xd}
  B.~C.~Barrois,
  ``Superconducting Quark Matter,''
  Nucl.\ Phys.\ B {\bf 129}, 390 (1977).

\bibitem{Bailin:1983bm}
  D.~Bailin and A.~Love,
  ``Superfluidity And Superconductivity In Relativistic Fermion Systems,''
  Phys.\ Rept.\  {\bf 107}, 325 (1984).

\bibitem{Alford:1997zt}
  M.~G.~Alford, K.~Rajagopal and F.~Wilczek,
  ``QCD at finite baryon density: Nucleon droplets and color
  superconductivity,''
  Phys.\ Lett.\ B {\bf 422}, 247 (1998)
  [arXiv:hep-ph/9711395].

\bibitem{Rapp:1997zu}
  R.~Rapp, T.~Schafer, E.~V.~Shuryak and M.~Velkovsky,
  ``Diquark Bose condensates in high density matter and instantons,''
  Phys.\ Rev.\ Lett.\  {\bf 81}, 53 (1998)
  [arXiv:hep-ph/9711396].

\bibitem{Alford:1998mk}
  M.~G.~Alford, K.~Rajagopal and F.~Wilczek,
  ``Color-flavor locking and chiral symmetry breaking in high density {QCD},''
  Nucl.\ Phys.\ B {\bf 537}, 443 (1999)
  [arXiv:hep-ph/9804403].

\bibitem{Deryagin:1992rw}
  D.~V.~Deryagin, D.~Y.~Grigoriev and V.~A.~Rubakov,
  ``Standing wave ground state in high density, zero temperature QCD at large
  N(c),''
  Int.\ J.\ Mod.\ Phys.\ A {\bf 7}, 659 (1992).

\bibitem{Shuster:1999tn}
  E.~Shuster and D.~T.~Son,
  ``On finite-density {QCD} at large N(c),''
  Nucl.\ Phys.\ B {\bf 573}, 434 (2000)
  [arXiv:hep-ph/9905448].




\bibitem{Shankar:1993pf}
  R.~Shankar,
  ``Renormalization group approach to interacting fermions,''
  Rev.\ Mod.\ Phys.\  {\bf 66}, 129 (1994).

\bibitem{Evans:1998ek}
  N.~J.~Evans, S.~D.~H.~Hsu and M.~Schwetz,
  ``An effective field theory approach to color superconductivity at high
  quark density,''
  Nucl.\ Phys.\ B {\bf 551}, 275 (1999)
  [arXiv:hep-ph/9808444].

\bibitem{Son:1998uk}
  D.~T.~Son,
  ``Superconductivity by long-range color magnetic interaction in  high-density
  quark matter,''
  Phys.\ Rev.\ D {\bf 59}, 094019 (1999)
  [arXiv:hep-ph/9812287].

\bibitem{Sannino:2004ix}
  F.~Sannino and K.~Tuominen,
  ``Tetracritical behavior in strongly interacting theories,''
  Phys.\ Rev.\ D {\bf 70}, 034019 (2004)
  [arXiv:hep-ph/0403175].

\bibitem{Mocsy:2003qw}
  A.~Mocsy, F.~Sannino and K.~Tuominen,
  ``Confinement versus chiral symmetry,''
  Phys.\ Rev.\ Lett.\  {\bf 92}, 182302 (2004)
  [arXiv:hep-ph/0308135]. 
  A.~Mocsy, F.~Sannino and K.~Tuominen,
  ``Critical behavior of non order-parameter fields,''
  Phys.\ Rev.\ Lett.\  {\bf 91}, 092004 (2003)
  [arXiv:hep-ph/0301229].

\bibitem{ILA} L.B.~Ioffe, D.~Lidsky, and
B.L.~Altshuler, Phys. Rev. Lett. {\bf 73}, 472 (1994);
B.L.~Altshuler, L.B.~Ioffe, A.J.~Millis Phys. Rev. B {\bf 50},
14048 (1994).


\end{thebibliography}
\end{document}